# $Cs_4Cr_7Te_{10}$: Interwoven Reconstructed Archimedean and Kagome Lattices with a Possible Phase Transition near 130 K

*Zhen Zhao[1], Ruwen Wang[1,2], Hua Zhang[1,2], Tong Liu[2], Haisen Liu[1], Guojing Hu[1], Ke Zhu[1], Senhao Lv[1], Gang Cao[1], Chenyu Bai[1,2], Hui Guo[1,2], Xiaoli Dong[1,2], Wu Zhou[2*], Haitao Yang[1,2*], and Hong-Jun Gao[1,2*]*

*[1] Beijing National Center for Condensed Matter Physics and Institute of Physics, Chinese Academy of Sciences, Beijing 100190, PR China*

*[2] School of Physical Sciences, University of Chinese Academy of Sciences, Beijing 100190, PR China*

[*]Correspondence to: hjgao@iphy.ac.cn, htyang@iphy.ac.cn, wuzhou@ucas.ac.cn

Chromium-based materials with complex lattice geometries provide an important platform for investigating correlated electronic and magnetic states. However, Cr-based compounds with unusual crystal geometries are still rarely reported. Here, we report a new Cr-based compound, $Cs_4Cr_7Te_{10}$, featuring interwoven Cr and Te sublattices that can be viewed as reconstructed networks derived from Archimedean 3.4.6.4 tiling and the kagome lattice, respectively. Transport measurements reveal the semiconducting nature in $Cs_4Cr_7Te_{10}$. Magnetization measurements show a weak anisotropy between *H*//*b* and *H*//*ac* planes, and uncover an anomaly near 130 K that is insensitive to the applied magnetic fields. Specific-heat measurements further confirm this transition, indicating its bulk thermodynamic nature. The associated entropy change is as small as 0.41 J $mol^{-1}$ $K^{-1}$, ruling out a structural phase transition and pointing to a possible electronic and/or magnetic phase transition. These results provide a new route for designing complex crystal geometries and exploring their associated emergent phenomena.

Chromium (Cr)-based materials have attracted extensive interest in condensed matter physics and materials science due to their unique 3d electronic configuration and moderate electron correlation effects. The partially filled d orbitals of Cr enable a wide variety of magnetic behaviors, including ferromagnetism, antiferromagnetism, and more complex spin orders. Meanwhile, the interplay between electron correlation and spin–orbit coupling in these systems provides a fertile platform for realizing novel topological quantum states. Both theoretical and experimental studies have demonstrated that Cr-based materials can host a range of intriguing physical properties, such as spin-polarized transport, topological semimetal states, and the quantum spin Hall effect. In low-dimensional systems, they are also capable of stabilizing two-dimensional magnetic order and supporting topological spin textures such as skyrmions, making them highly promising for spintronic and quantum device applications. Therefore, exploring new Cr-based compounds is of great importance for understanding the coupling between magnetism and topology.

A wide range of exotic phenomena has been reported or predicted in these systems. For instance, $HgCr_2Se_4$ has been identified as ferromagnetic half-metals and further proposed as Chern semimetals hosting topologically protected Weyl fermions[1,2]. Layered and van der Waals Cr-based materials, including $CrI_3$ and $Cr_2Ge_2Te_6$, exhibit intrinsic two-dimensional magnetism down to the monolayer limit, opening new avenues for low-dimensional spintronic devices[3,4]. Meanwhile, Cr–Te compounds have demonstrated robust room-temperature magnetism and the ability to host nontrivial spin textures such as skyrmions, accompanied by pronounced topological Hall effects[5,6]. In addition, kagome-lattice systems such as $CsCr_3Sb_5$ provide a fertile ground for studying the coexistence of flat bands, Dirac dispersions, and strong electronic correlations[7,8]. Beyond magnetism, certain Cr compounds such as $K_2Cr_3As_3$ also display spin-triplet superconductivity, highlighting the competition and coexistence between different quantum orders[9–11]. Collectively, these findings demonstrate that Cr-based materials host a rich variety of emergent phenomena, making them promising candidates for both fundamental studies and future quantum technologies.

Following the discovery of $Cs_3V_9Te_{13}$[12], we sought to further expand this material family through partial atomic substitution. Beyond replacing the Cs atoms with other alkali metals, such as Rb, which led to the successful synthesis of $Rb_3V_9Te_{13}$ with semiconducting behavior, we also explored substitution at the transition-metal site. Motivated by the rich magnetic properties of Cr and its close proximity to V in the periodic table, we attempted to replace V with Cr. This effort led to the successful synthesis of a new Cr-based compound, $Cs_4Cr_7Te_{10}$. This material exhibits semiconducting transport behavior and a possible phase transition near 130 K without detectable long-range magnetic orders,. More importantly, its crystal structure consists of two interwoven sublattices: a Cr network reconstructed from Archimedean 3.4.6.4 tiling and a Te network reconstructed from the kagome lattice, providing a new structural platform for exploring the relationship between the lattice reconstruction and emergent physical properties.

Single crystals of $Cs_4Cr_7Te_{10}$ were synthesized by a self-flux method similar to our recent synthesis method for $Cs_3V_9Te_{13}$[12]. The crystal structure was determined by the single-crystal X-ray diffraction. $Cs_4Cr_7Te_{10}$ crystallizes in a monoclinic structure with the space group C2/m (No. 12). The refined lattice parameters are $a$ = 10.2883(18) Å, $b$ = 17.247(3) Å, and $c$ = 7.7082(12) Å, with $\alpha$ = 90°, $\beta$ = 97.589(5) °, and $\gamma$ = 90°. Further crystallographic details, including structural parameters and atomic coordinates, are provided in Tables S1–S4. As illustrated in Fig. 1a, the Cr atoms form a complex sublattice built from triangle- and quadrilateral-based motifs. For clarity, this Cr framework can be understood as a reconstructed lattice derived from the Archimedean 3.4.6.4 tiling (Fig. S1), formed through the bond breaking and subsequent lattice sliding. Te atoms, shown in Fig. 1b, form another distinct sublattice composed of triangles and octagons. Despite this modified connectivity, the Te network retains a clear geometric relationship to the kagome lattice and can therefore be described as a reconstructed kagome lattice generated through a similar process of the bond breaking and lattice sliding from a parent kagome network (Fig. S2). Notably, the crystal structure of $Cs_4Cr_7Te_{10}$ consists of two interwoven reconstructed sublattices derived from distinct parent geometries but sharing a common reconstruction mechanism. Accordingly, the Cr and Te sublattices can be described as a split-intercalated Archimedean lattice and a split-intercalated kagome lattice, respectively. To the best of our knowledge, this type of atomic sublattice has not been reported previously. The lattice geometry of a crystal plays a central role in governing electronic correlations. In most cases, atomic networks tend to adopt highly symmetric motifs, such as honeycomb, kagome, or Lieb lattices, which are generally associated with an enhanced structural stability. In $Cs4Cr_7Te_{10}$, however, the system stabilizes an unusually complex and distorted lattice geometry with the reduced symmetry. The emergence of such a structurally intricate framework expands the range of accessible crystal architectures, provides a new avenue for materials design and exploration, and may offer a novel structural platform for investigating the geometrical frustration and related emergent phenomena.

An optical image of a typical crystal is shown in the inset of Fig. 2a. The crystal naturally forms a hexagonal prismatic shape and can be mechanically exfoliated along one pair of side facets. The XRD pattern collected from the exfoliated surface is shown in Fig. 2a. The diffraction peaks can be indexed to the *(0l0)* family of planes, indicating that the exposed surface corresponds to the *ac* plane. Energy-dispersive spectroscopy (EDS) result, as shown in Fig. 2b, yields a semi-quantitative composition of Cs: Cr: Te = 3.8: 7: 9.6, which is consistent with the stoichiometric ratio of $Cs_4Cr_7Te_{10}$. To further verify the novel structure, atomic-resolution STEM characterization was carried out. As shown in Fig. 2c-d, the experimental HAADF-STEM images clearly resolve the layered structure, and the projected atomic models match well with the experimental images. The consistency between the projected structural model and the experimental atomic image further supports the validity of the proposed crystal structure. Moreover, the FFT pattern agrees well with the simulated electron diffraction pattern along the [001] zone axis, confirming the viewing direction. No discernible defects, impurities, or other structural inhomogeneities are observed, indicating the high quality of the sample. Therefore, these combined results establish $Cs_4Cr_7Te_{10}$ as a newly synthesized compound with a previously unreported crystal structure. It is worth noting that the sample is highly unstable and rapidly degrades upon exposure to air, much similar to that observed in our previously-reported kagome metal $CsTi_3Bi_5$[13–15].

After the synthesis of the $Cs_4Cr_7Te_{10}$ single crystal, we carried out electrical transport measurements as a first characterization. As shown in Fig. S3, the temperature-dependent resistivity, $\rho(T)$, indicates that $Cs_4Cr_7Te_{10}$ exhibits semiconducting behavior, with the resistivity increasing upon cooling. The resistivity increases markedly from 1.7 Ω·m at room temperature to 832 Ω·m at 10 K. To rule out the possibility that the observed semiconducting behavior is extrinsic and caused by the sample degradation, we repeated the measurement on several different crystals and obtained consistent results, thereby confirming the intrinsic semiconducting nature of $Cs_4Cr_7Te_{10}$.

As the presence of Cr naturally raises the possibility of magnetic ordering, we performed magnetization measurements with the magnetic field applied both along the *b* axis (H//b) and parallel to the *ac* plane (*H*//*ac* plane). The magnetic properties of $Cs_4Cr_7Te_{10}$ were first investigated by temperature-dependent magnetic susceptibility measurements. As shown in Fig. 3a, the magnetic susceptibility measured under $H$ // $b$ at 0.1 T shows no pronounced bifurcation between the zero-field-cooled (ZFC) and field-cooled (FC) curves, indicating the absence of long-range ferromagnetic ordering. Overall, the susceptibility increases upon cooling and exhibits predominantly paramagnetic-like behavior over the measured temperature range, with a kink appearing near 130 K. An enlarged view in Fig. 3b highlights this anomaly at $T = 130$ K. To further investigate its field dependence, additional measurements were carried out under applied fields of 1 T and 5 T. The transition temperature shows no apparent shift with increasing magnetic fields, indicating that the anomaly is robust against the applied magnetic fields. This behavior suggests that the feature is intrinsic in origin and may be associated with a phase transition, such as a density-wave instability[16].

For comparison, the in-plane magnetic susceptibility measured with *H // ac* plane is shown in Fig. 3c-d. Similar to the out-of-plane data, no obvious ZFC–FC splitting is observed, further supporting the absence of long-range ferromagnetic ordering. The susceptibility increases monotonically upon cooling, characteristic of a paramagnetic-like response. A weak anomaly remains discernible near 130 K in the in-plane data, confirming that this feature is intrinsic to the bulk material. However, compared with the H // *b* configuration, the anomaly is significantly weaker and appears only as a subtle change in slope across 130 K, indicating a weak magnetic anisotropy in $Cs_4Cr_7Te_{10}$. To gain the additional insight, we performed isothermal magnetization measurements, as shown in Fig. S4. A nearly linear field dependence is observed over the entire magnetic-field range, without detectable hysteresis loops, ruling out the existence of ferromagnetism. This result is consistent with the temperature-dependent magnetization measurements. Overall, these results suggest that $Cs_4Cr_7Te_{10}$ does not host the long-range magnetic ordering down to low temperatures. Instead, the presence of a weak anomaly around 130 K may point to a possible phase transition, potentially associated with subtle changes in the electronic structure or short-range magnetic correlations. Interestingly, this behavior bears resemblance to that observed in our recently reported heavy-fermion material $Cs_3V_9Te_{13}$[12].

To further clarify the anomaly observed in the magnetic measurements, specific-heat measurements were carried out on $Cs_4Cr_7Te_{10}$, as shown in Fig. 4. The specific heat varies smoothly with temperatures over the whole measured range and does not exhibit a pronounced lambda-type anomaly, indicating the absence of a conventional bulk long-range phase transition. However, a weak feature is clearly resolved near 130 K, in good agreement with the anomaly observed in the magnetic susceptibility. As further shown in the inset, the specific heat data collected under 0 T and 5 T nearly overlap, without noticeable field-induced changes. This result suggests that the transition is insensitive to the applied magnetic fields, which is consistent with the magnetic measurements.

The Debye temperature was estimated from the linear fitting of the low-temperature specific-heat data in the *C/T* versus $T^2$ plot (Fig. 4b). The fitted $\beta$ value is 0.0415 J mol$^{-1}$ K$^{-4}$, corresponding to a Debye temperature of 99 K. By subtracting the background, the entropy change associated with the anomaly at 130 K was determined to be 0.41 J mol$^{-1}$ K$^{-1}$, as shown in Fig. 4c and 4d. Such a small entropy change is unlikely to be associated with a structural phase transition, in agreement with the temperature-dependent XRD results in Fig. S5.

In summary, we have synthesized a new layered Cr-based compound, $Cs_4Cr_7Te_{10}$, which hosts an unusual lattice architecture composed of interwoven reconstructed Cr and Te sublattices derived from Archimedean 3.4.6.4 tiling and the kagome lattice, respectively. Structural characterization reveals an architecture with intertwined atomic motifs, suggesting a potential platform for geometrical frustration and correlated electronic behavior. Transport measurements reveal the semiconducting nature over a wide temperature range. Magnetic susceptibility shows an overall paramagnetic-like response without the long-range ferromagnetic ordering. A weak anomaly is observed near 130 K, which is insensitive to applied magnetic fields. Specific heat results further confirm the existence of a bulk transition at 130 K, although the absence of a pronounced lambda-type anomaly and the small associated entropy change indicate that this transition is not of the conventional long-range magnetic or structural origin. Taken together, these results present that $Cs_4Cr_7Te_{10}$ exhibits a non-structural transition near 130 K, pointing to a possible density-wave transition. Therefore, $Cs_4Cr_7Te_{10}$ provides a new platform for exploring the interplay between lattice complexity, electronic correlations, and emergent phenomena.

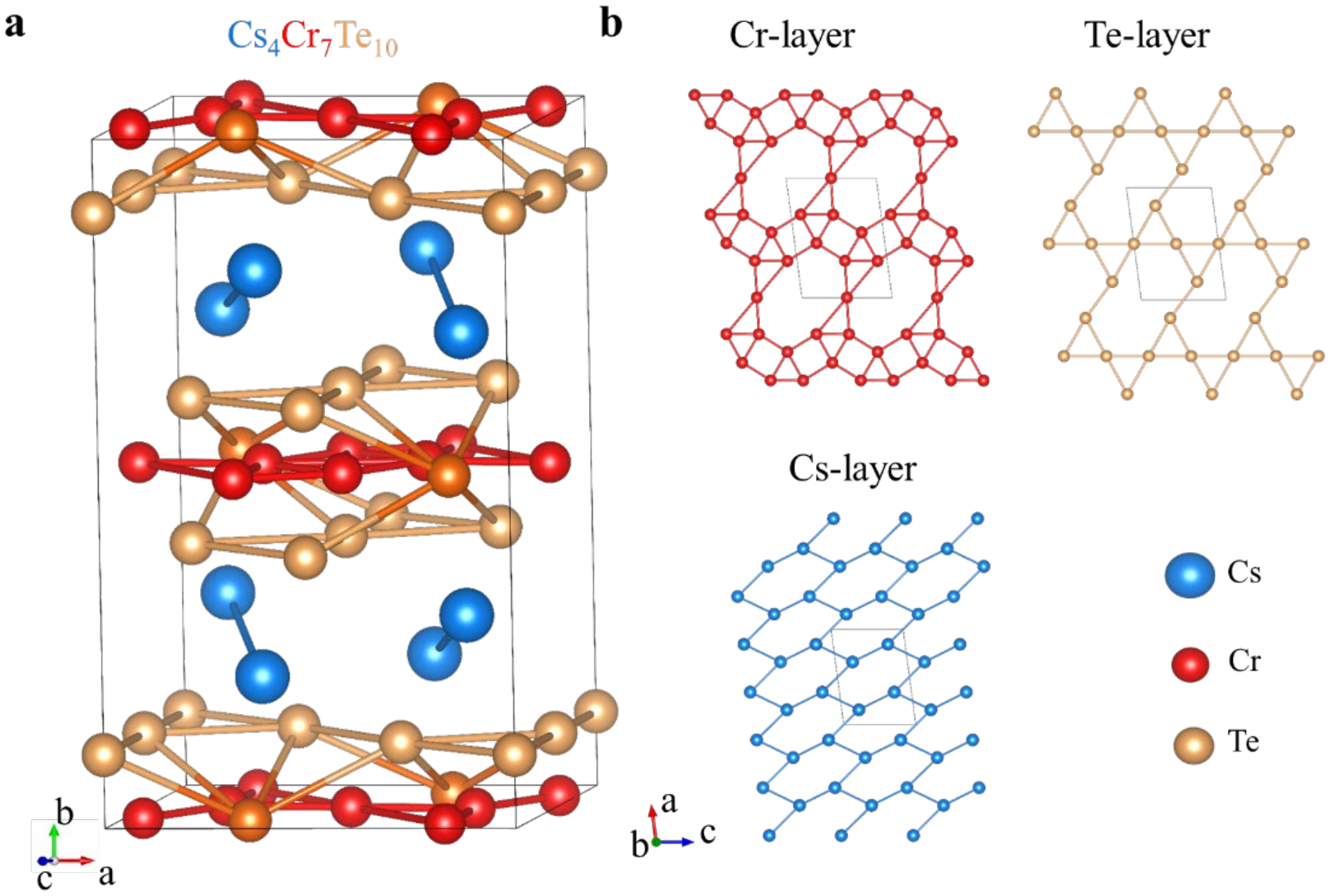

**Figure 1. crystal structural motif of $Cs_4Cr_7Te_{10}$.**
(a-b) Schematic crystal structure of $Cs_4Cr_7Te_{10}$, with Cs atoms shown in blue, Cr atoms in red, and Te atoms in yellow. The solid lines denote the unit cell.

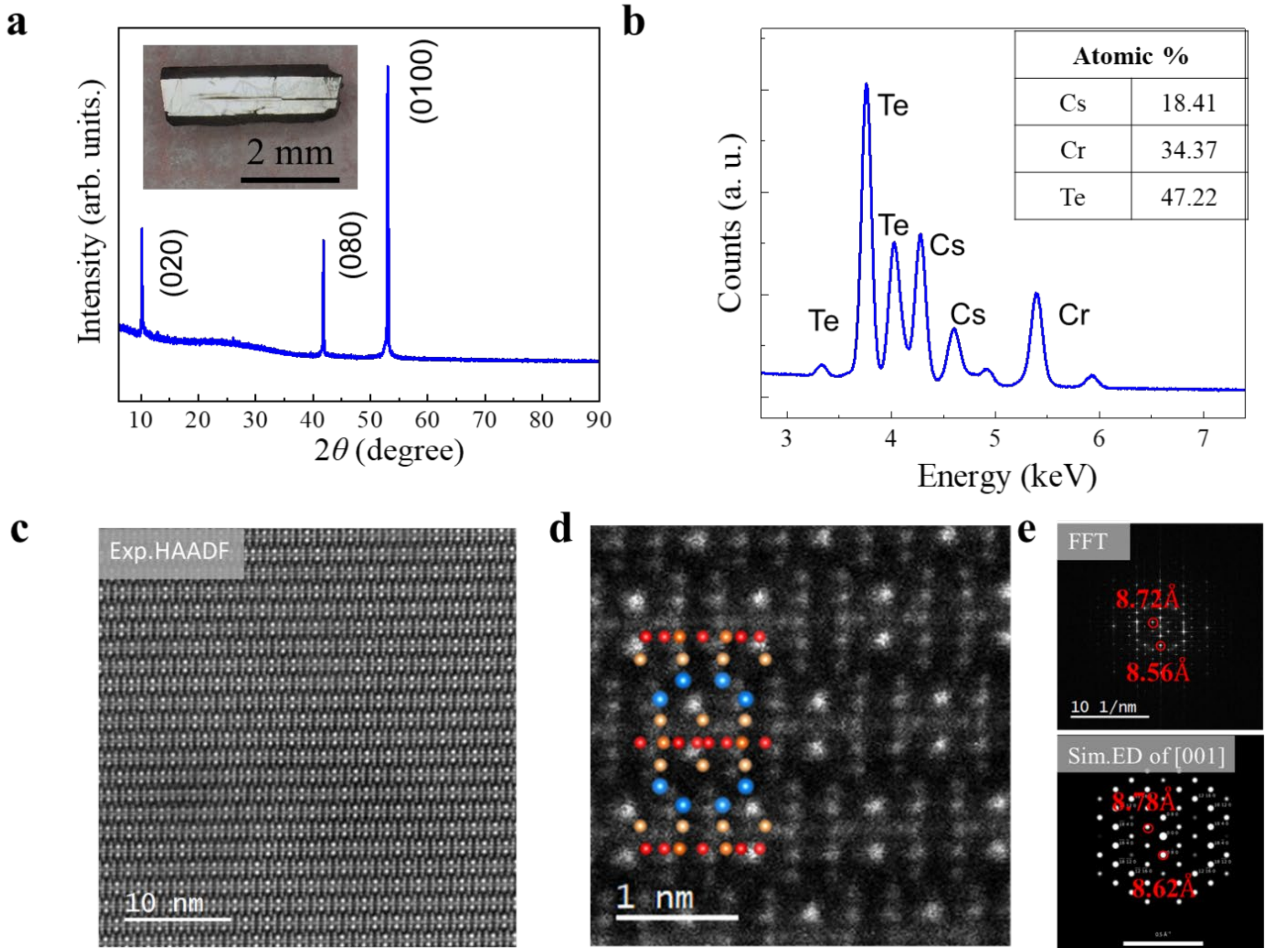

**Figure 2. Structural characterization and compositional analysis of the $Cs_4Cr_7Te_{10}$ crystal.**

**(a)** Single-crystal X-ray diffraction pattern of $Cs_4Cr_7Te_{10}$, showing only (0l0) reflections consistent with its layered structure. The inset shows an optical image of a representative crystal. **(b)** EDS spectrum and atomic percentage analysis of $Cs_4Cr_7Te_{10}$, yielding a composition close to the nominal stoichiometry with Cs: Cr: Te = 3.8: 7: 9.6. **(c-d)** Atomic-resolution HAADF-STEM image viewed, with the corresponding structural model superimposed, where Cs, Cr, and Te atoms are represented by blue, red, and yellow spheres, respectively. (e) Fast Fourier transform (FFT) pattern derived from the STEM image, together with the simulated electron diffraction pattern along the [001] direction.

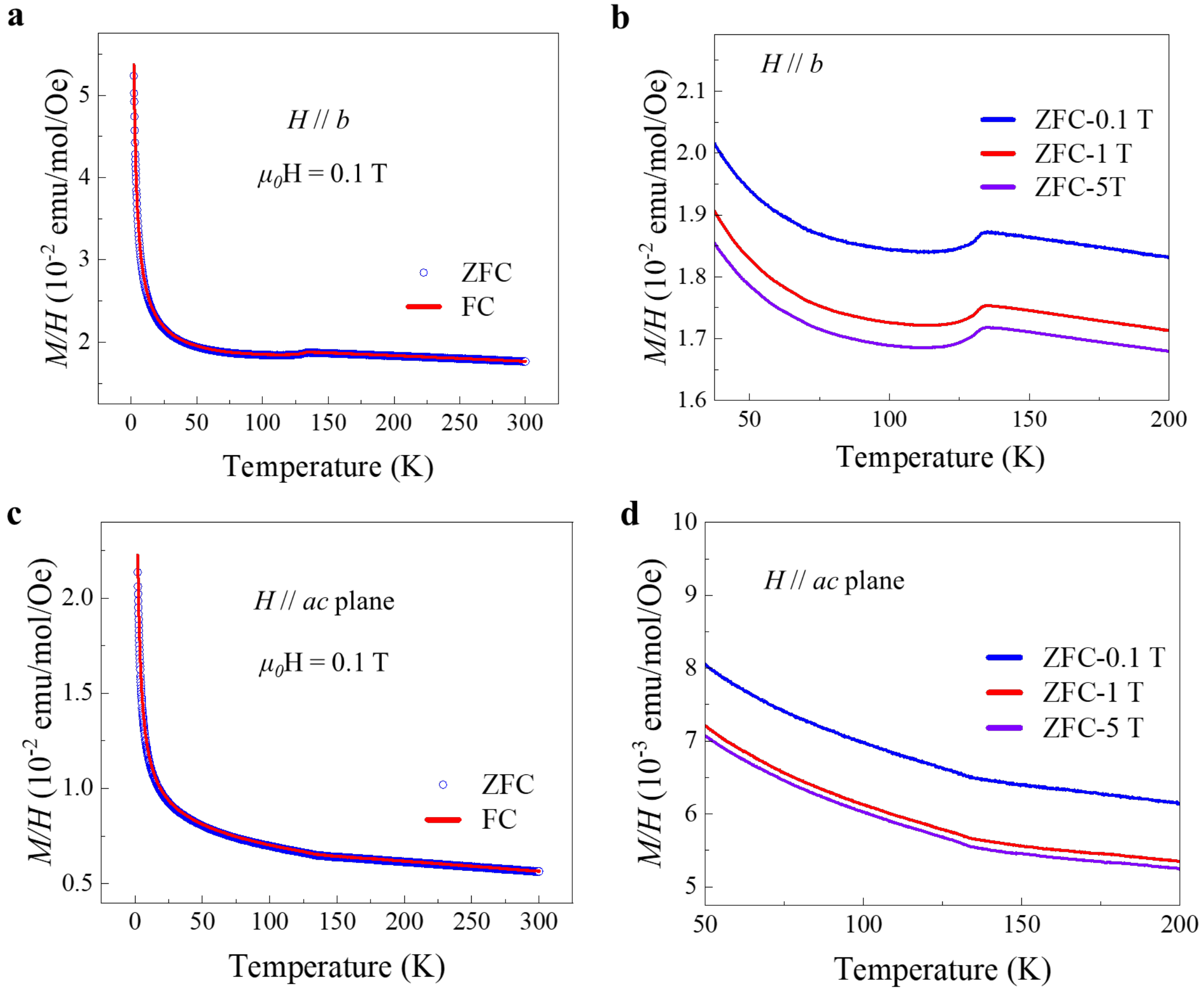

**Figure 3. Temperature-dependent magnetic susceptibility of $Cs_4Cr_7Te_{10}$.**

(a) Magnetic susceptibility (M/H) measured under $\mu_0H = 0.1$ T with the magnetic field applied along the *b* axis. (b) Enlarged view of M/H under different magnetic fields of 0.1 T, 1 T, and 5 T, revealing a weak anomaly near 130 K that is insensitive to the applied magnetic fields. (c) Magnetic susceptibility measured under $\mu_0H = 0.1$ T with the magnetic field applied in the *ac* plane. (d) Enlarged view of M/H under different magnetic fields of 0.1 T, 1 T, and 5 T, also revealing a weak anomaly near 130 K.

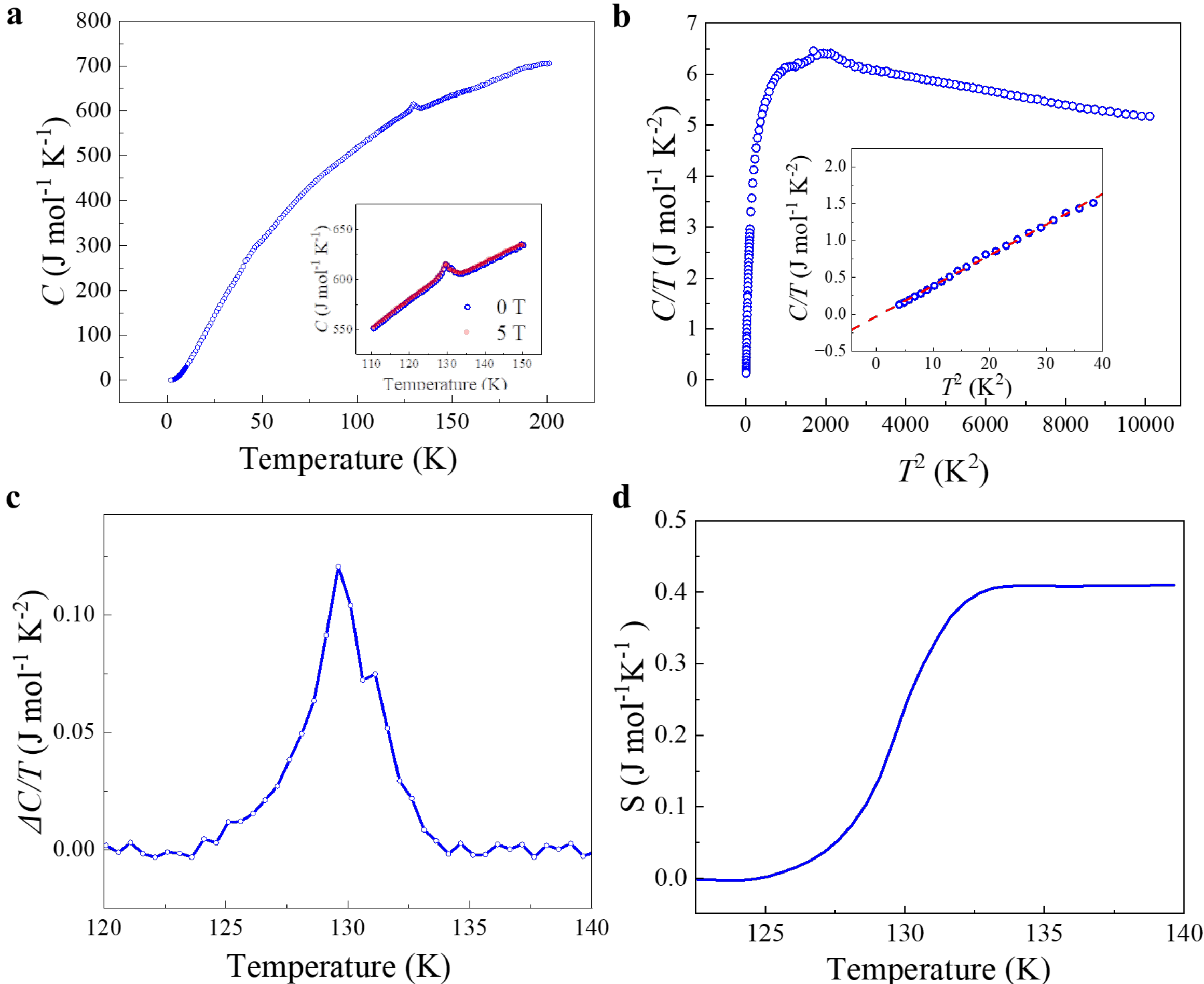

**Figure 4. Specific heat measurement of $Cs_4Cr_7Te_{10}$.**

(a) Temperature dependence of specific heat $C(T)$, with a weak feature near 130 K; inset compares data measured under 0 T and 5 T. (b) Plot of $C/T$ versus $T^2$ at low temperatures. The dashed line represents a linear fit in the 2–6 K range using $C/T = \gamma + \beta T^2$, yielding $\beta$ = 41.5 mJ mol$^{-1}$ K$^{-4}$. (c) Temperature dependence of $\Delta C/T$ after background subtraction, highlighting a weak anomaly around 130 K. (d) Temperature-dependent entropy changes S(T), indicating a small entropy release (~0.41 J mol$^{-1}$ K$^{-1}$) associated with the transition.

## Supporting Information

**Table S1. Sample and crystal data for $Cs_4Cr_7Te_{10}$.**

| Chemical formula | $Cs_4Cr_7Te_{10}$ |
|---|---|
| Formula weight | 2171.64 g/mol |
| Temperature | 302(2) K |
| Wavelength | 0.71073 Å |
| Crystal size | 0.024 x 0.169 x 0.594 mm |
| Crystal system | monoclinic |
| Space group | C 1 2/m 1 |
| a (Å) | 10.2883(18) |
| b (Å) | 17.247(3) |
| c (Å) | 7.7082(12) |
| α (°) | 90 |
| β (°) | 97.589(5) |
| γ (°) | 90 |
| Volume ($Å^3$) | 1355.8(4) |
| Z | 2 |
| Density (calculated) | 5.319 $g/cm^3$ |
| Absorption coefficient | 18.525 $mm^{-1}$ |
| F(000) | 1816 |

**Table S2. Data collection and structure refinement for $Cs_4Cr_7Te_{10}$.**

| Parameter | Value |
|---|---|
| Theta range for data collection | 2.32 to 28.32° |
| Index ranges | -13<=h<=13, -22<=k<=22, -10<=l<=10 |
| Reflections collected | 8412 |
| Independent reflections | 1742 [R(int)=0.0520] |
| Max. and min. transmission | 0.6650 and 0.0320 |
| Refinement method | Full-matrix least-squares on $F^2$ |
| Refinement program | SHELXL-2019/1 |
| Function minimized | $\Sigma\, w(Fo^2 - Fc^2)^2$ |
| Data / restraints / parameters | 1742 / 0 / 57 |
| Goodness-of-fit on $F^2$ | 1.068 |
| Final R indices (I>2σ(I)) | R1=0.0550, wR2=0.1678 |
| Final R indices (all data) | R1=0.0589, wR2=0.1737 |
| Weighting scheme | $w=1/[\sigma^2(Fo^2)+(0.1229P)^2+38.4958P]$, $P=(Fo^2+2Fc^2)/3$ |
| Largest diff. peak and hole | 3.572 and -2.500 eÅ$^{-3}$ |
| R.M.S. deviation from mean | 0.668 eÅ$^{-3}$ |

**Table S3. Atomic coordinates and equivalent isotropic atomic displacement parameters (Å$^2$) for $Cs_4Cr_7Te_{10}$.**

| Atom | x/a | y/b | z/c | U(eq) |
|---|---|---|---|---|
| Te1 | 0.500000 | 0.60692(5) | 0.500000 | 0.0182(3) |
| Te2 | 0.500000 | 0.39308(5) | 0.000000 | 0.0180(3) |
| Cs1 | 0.15760(9) | 0.29443(6) | 0.23771(15) | 0.0494(3) |
| Te03 | 0.16151(6) | 0.39460(4) | 0.68623(9) | 0.0196(2) |
| Te04 | 0.18561(9) | 0.500000 | 0.19161(13) | 0.0257(3) |
| Cr1 | 0.3055(2) | 0.500000 | 0.8921(3) | 0.0167(5) |
| Cr07 | 0.3073(2) | 0.500000 | 0.5317(3) | 0.0173(5) |
| Cr08 | 0.4552(2) | 0.500000 | 0.2417(3) | 0.0171(5) |
| Cr09 | 0.000000 | 0.500000 | 0.500000 | 0.0346(10) |

**Table S4. Anisotropic atomic displacement parameters (Å$^2$) for $Cs_4Cr_7Te_{10}$.**

| Atom | U11 | U22 | U33 | U23 | U13 | U12 |
|---|---|---|---|---|---|---|
| Te1 | 0.0183(5) | 0.0218(5) | 0.0148(5) | 0.000000 | 0.0028(4) | 0.000000 |
| Te2 | 0.0193(5) | 0.0191(5) | 0.0150(5) | 0.000000 | -0.0002(3) | 0.000000 |
| Cs1 | 0.0458(6) | 0.0386(5) | 0.0610(7) | -0.0161(4) | -0.0032(5) | 0.0098(4) |
| Te03 | 0.0162(4) | 0.0212(4) | 0.0208(4) | 0.0012(2) | -0.0002(3) | -0.0013(2) |
| Te04 | 0.0172(5) | 0.0440(7) | 0.0163(5) | 0.000000 | 0.0030(4) | 0.000000 |
| Cr1 | 0.0143(9) | 0.0213(10) | 0.0140(10) | 0.000000 | -0.0001(8) | 0.000000 |
| Cr07 | 0.0142(9) | 0.0245(11) | 0.0137(10) | 0.000000 | 0.0034(8) | 0.000000 |
| Cr08 | 0.0170(10) | 0.0223(11) | 0.0120(10) | 0.000000 | 0.0021(8) | 0.000000 |
| Cr09 | 0.0290(18) | 0.0204(17) | 0.046(2) | 0.000000 | -0.0245(17) | 0.000000 |

The anisotropic atomic displacement factor exponent takes the form: $-2\pi^2[h^2a^{*2}U_{11} + ... + 2hka^*b^*U_{12}]$